\documentclass[10pt,twocolumn]{article}

\usepackage{graphicx}
\usepackage{multicol}
\usepackage{amssymb}
\usepackage{amsmath}
\usepackage{booktabs,multirow}
\usepackage{color}
\usepackage{url}
\usepackage{cite}
\usepackage{scrextend}
\usepackage{authblk}

\newcommand{\tA}{\mathcal{A}}

\newcommand{\tD}{\mathcal{D}}

\newcommand{\di}{\mathrm{diag}}
\newcommand{\tG}{\mathcal{G}}

\newcommand{\tE}{\mathcal{E}}
\newcommand{\tL}{\mathcal{L}}

\newcommand{\tN}{\mathcal{N}}

\newcommand{\bT}{\mathbb{T}}
\newcommand{\bR}{\mathbb{R}}

\newtheorem{rem}{Remark}

\DeclareMathOperator*{\argmin}{argmin}

\date{}

\title{Distributed Peer-to-Peer Energy Trading for Residential Fuel Cell Combined Heat and Power Systems}

\author[1]{Dinh Hoa Nguyen $^{\ast}$} 
\author[2]{Tatsumi Ishihara} 
\affil[1]{International Institute for Carbon-Neutral Energy Research (WPI-I$^2$CNER), and the Institute of Mathematics for Industry (IMI), Kyushu University, Fukuoka 819-0395, Japan. E-mail: hoa.nd@i2cner.kyushu-u.ac.jp}
\affil[2]{International Institute for Carbon-Neutral Energy Research (WPI-I$^2$CNER), and Department of Applied Chemistry, Faculty of Engineering, Kyushu University, Fukuoka 819-0395, Japan. E-mail: ishihara@cstf.kyushu-u.ac.jp.}

\begin{document}

\maketitle

\begin{abstract}
This paper studies the optimal energy management in a group of dwellings having micro fuel cell combined heat and power systems. To increase the self-sufficiency and resilience of such local community, a peer-to-peer energy trading system between dwellings is proposed in which output powers from fuel cells working under their rated powers can be sold to those already reach their rated outputs but still lack powers. The arising optimization problem from this optimal peer-to-peer energy trading system is non-convex due to the nonlinear dependence of power and heat efficiencies on fuel cell output power. Therefore, a linearization method is proposed to convexify the problem. Consequently, a distributed ADMM approach is introduced to solve the convexified optimization problem in parallel at each dwelling. A case study for a group of six dwellings based on  realistic electric consumption data is then presented to demonstrate the proposed approach performance and positive impacts of the P2P energy trading system. More specifically, the proposed distributed ADMM approach is reasonably fast in convergence and is scalable well with system size. In addition, P2P electricity trading system helps operate fuel cells at a higher efficiency and increase the self-sufficiency of such dwellings.   
\end{abstract}


\section*{Nomenclature}
\begin{labeling}{Lowerupper bounds}
\item[P2P] Peer to peer.
\item[DES] Dwelling electricity system.
\item[DER] Distributed energy resource.
\item[ADMM] Alternating direction method of multipliers. 
\item[FC, SOFC] Fuel cell, solid oxide fuel cell.
\item[CHP] Combined heat and power.
\item[$\eta_{i,e}$, $\eta_{i,g2h}$] Efficiency of FC electricity output and  fuel-to-hydrogen processing at dwelling $i$.
\item[$\eta_{i,hr}$] Efficiency of FC heat recovery at dwelling $i$.
\item[$\xi_{e}$] Conversion factor for electricity [MJ/kWh].
\item[$HT_{i,in}$] Hot water charged to storage tank at dwelling $i$ [l]. 
\item[$T_{i,ht}$, $T_{i,cn}$]  Water temperature from hot tank and from city water network at dwelling $i$ [$^{\circ}$C]. 
\item[$t$, $\Delta t$, $\bT$]  Time step index, time step length, and maximum time in the considered period. 
\item[$n$]  Number of dwellings. 
\item[$p_{g}$]  City gas price [Y/MJ]. 
\item[$q_{w}$]  Specific heat of water [MJ/l$^{\circ}$C]. 
\item [$E_{i,fc}$] Natural gas energy consumption by FC and backup boiler, at dwelling $i$ [MJ].
\item [$P_{i,fc}$] Power generated by FC at dwelling $i$ [kW].
\item [$P_{i,fc}^{\min}, \, P_{i,fc}^{\max}$] Lower and upper bounds on generated power of FC at dwelling $i$ [kW].
\item [$P_{i,dem}$] Electric demand of dwelling $i$ [kW].
\item [$P_{i,grid}$] Power bought from grid by dwelling $i$ [kW].
\item [$P_{ij}$, $P_{i,tr}$] P2P traded electricity between dwellings $i$ and $j$, and total P2P traded electricity of dwelling $i$ [kW].
\item [$P_{i,tr}^{\min}$, $P_{i,tr}^{\max}$] Lower and upper bounds of total P2P traded electricity of dwelling $i$ [kW].
\item[$\tG, \, \tE, \, \tA, \, \tD, \, \tL$] P2P trading graph, its edge set, its adjacency, degree, and Laplacian matrices.
\end{labeling}

\section{Introduction}

The use of FC-CHP for co-generation systems in residential and commercial buildings was shown to be promising, where many advantages are obtained \cite{Arsalis19,DOE-CHP,Ellamla15}, despite their relatively high system cost. 
Those advantages include high overall (electricity and thermal) system efficiency, low emissions of pollutants, and potential decrease of electric bills. Therefore, FC-CHP co-generation system is an attractive type of DERs installed at dwellings for local generation and consumption. As such, dwellings are less dependent on the bulk grid, especially when power outages occur. Moreover, harmful effects to the grid voltage and frequency stability caused by reverse power flows from renewable and DERs can be reduced. 

To enhance local generation and consumption, new energy trading mechanisms for renewable and DERs have recently been extensively investigated. One of the promising candidates is the so-called P2P energy market \cite{Baez-Gonzalez18,Sousa19,Tushar18,Tushar20} which offers remarkable features in addition to what were mentioned above. 
First, P2P trading platforms are usually decentralized and localized systems, which are very suitable for integrating DERs and give much more flexibility for prosumers to handle their energy balance and benefit. Second, energy losses are reduced in P2P systems because energy is exchanged within short distances. As a result, investment cost is lower. Third, equipped with distributed ledger technologies such as block-chain, the security and privacy in P2P markets are much better than that in the conventional bulk energy grids \cite{Tushar18}. Next, P2P trading promotes new businesses since different models and market scales can be performed under P2P energy system concept, e.g. federated plans \cite{MorstynP2P18}, full P2P, community-based, or their hybrid combination \cite{Sousa19,Moret19}. In summary, P2P systems will serve as an important block to transform the current top-down, centralized energy networks into bottom-up, decentralized ones.  All of these motivate us to propose a P2P electricity trading system in the current research for local communities having FC-CHP units.  

A characteristic making P2P energy market different from other energy markets is on the direct energy trading between each prosumer/peer/agent with another communicated prosumer/peer/agent. Hence, a constraint on power balance is enforced to each pair of communicated peers, unlike the only one balance constraint for the total generated and consumed powers of all producers and consumers in other markets. To cope with these individual power balance constraints, several different P2P trading schemes have been introduced, e.g. game theory based \cite{Tushar19,Tushar18,Moret19,Cadre20,SCui20,SCui19}, bilateral contracts \cite{Sorin19,Baroche19,MorstynP2P19b,Khorasany19}, multi-class energy management \cite{MorstynP2P19a}, continuous double auction \cite{GuerreroA19}, distribution optimal power flow \cite{GuerreroA19}, supply-demand ratio based pricing \cite{NLiu17}, mixed performance indexes \cite{Werth18}, Lyapunov optimization \cite{NLiu18}, etc. 

The existing works on P2P energy systems usually assume a successful energy transaction between any pair of peers/agents at any time step, and often consider fixed roles of buyers/sellers for all time steps or do not restrict a peer to be a buyer or seller at a time step. However, the former assumption is not always held in realistic contexts, because some peer might not agree with the energy price or energy amount to be traded, and hence will not successfully trade. The latter assumption is not suitable for time-varying behaviors of prosumers who can both sell or buy energy, but act explicitly as a buyer or a seller at one time step, based on their predicted energy demands and limited or no storage capacity (to reduce investment cost). 

On the other hand, the optimal energy management problem for multiple FC-CHP systems was investigated in several works, e.g.  \cite{Wakui10,Aki16,Aki18,Tran18}. In those studies, they assumed that hot water can be exchanged between nearby dwellings. However, several issues arise from that assumption. First, an insulated piping network together with a pumping and controlling system must be constructed for such hot water exchange, which can significantly increase system cost. 
Second, FC output hot water temperature are usually less than 100$^{\circ}$C (e.g. 65$^{\circ}$C as in \cite{JLPGA}), which is lower when reaching another dwelling through the piping network, and hence could reduce its efficient use. Another critical assumption in the studies \cite{Wakui10,Aki16,Aki18} was that all FCs equally share the total energy demand from all dwellings. This is impractical because of: (i) different investment costs from different FC manufacturers; (ii) dissimilar energy demands between dwellings. 
Further, each dwelling owner always wants to control and manage its system in private, which does not allow such load sharing scheme. Last, any energy lack or excess of FC-CHP systems is bought from or sold to the bulk grid, which causes problems to the grid as discussed above.  
Energy management for grid-connected microgrids having CHP and PV prosumers has also been investigated using the Stackelberg game, e.g. \cite{LMa16}. Nevertheless, all the trading of prosumers were made with the microgrid operator, not between themselves.

To overcome the aforementioned drawbacks, this paper proposes an optimal gas-electricity management approach for FC-CHP-equipped dwellings participating in a P2P electricity trading system. 
Dwellings aim to minimize their energy costs from gas usage and from possible P2P electricity trading, where each of them behaves as a buyer or seller at one time step, but not both, and its role is changed depending on its predicted electric demand at the next time step. This results in a dynamic optimization problem with time-varying structure. Hence, it cannot be solved for all time steps at once like that for non-P2P market in \cite{Tran18}, but at each time step. Moreover, this mathematical programming is non-convex due to the nonlinearity of the FC power efficiency curve. Our first contribution is then to explicitly take into account the time-varying feature of P2P energy trading markets for real-time optimal energy management. Next, our second contribution is to propose a convexification method by linearizing the FC power efficiency curve, under which the optimal energy management problem can be more efficiently solved with much lower computational efforts and costs. Our last contribution is a distributed and parallel method to solve the convexified problem by each dwelling, based on the ADMM, which converges reasonably fast in tens of iterations.   

The rest of this paper is organized as follows. Section \ref{sysmod} introduces the energy system model of dwellings used in this research. Then a novel energy management strategy including a P2P energy trading mechanism will be presented in Section \ref{proposed-approach}. Next, a case study are given in Section \ref{cases} to illustrate the performance and benefits of our proposed approach. Finally, conclusions are provided in Section \ref{sum}.

	\begin{figure*}[htpb!]
		\centering
		\includegraphics[scale=0.45]{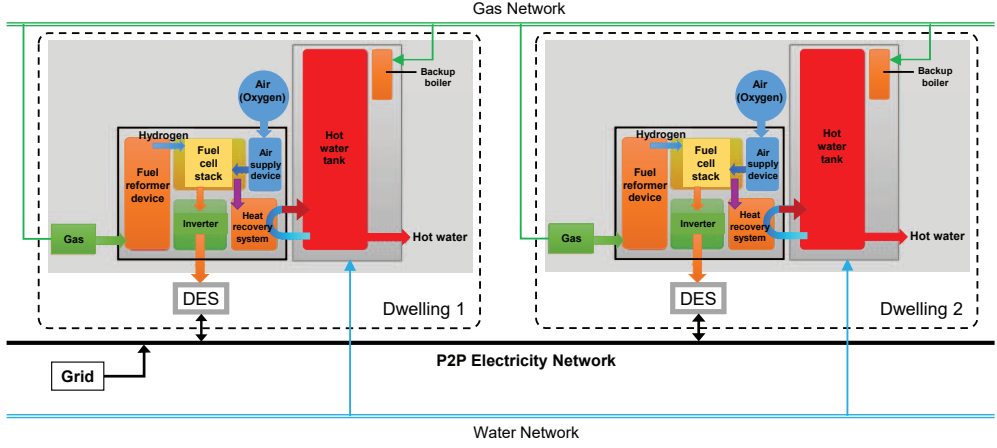}
		\caption{Illustration for the proposed P2P electricity trading system for multiple dwellings equipped with FC-CHP units.} 
		\label{EGWN}
	\end{figure*}	

\section{System Model}
\label{sysmod}

The overall system configuration of dwellings with FC-CHP units is shown in Figure \ref{EGWN}. 
Details on the mathematical models of system components are presented below. 

\subsection{Fuel Cell}
\label{fc}

This research studies a SOFC system using gas (whether natural gas or liquefied petroleum gas) 
as the input fuel. The gas energy consumed by the FC at dwelling $i$ is computed by
\begin{align}
\label{g-1}
E_{i,fc}(t) = \dfrac{P_{i,fc}(t)\Delta t \xi_e}{\eta_{i,e}(t)\eta_{i,g2h}}
\end{align}
Hence, the cost of gas usage at dwelling $i$ is   
\begin{align}
\label{g-cost}
C_{i,g}(t) = p_g(t) E_{i,fc}(t) = p_g(t) \dfrac{P_{i,fc}(t)\Delta t \xi_e}{\eta_{i,e}(t)\eta_{i,g2h}}
\end{align}
SOFC-CHP system is the main source of power generation for dwellings, which continuously runs throughout the day \cite{Ozawa18}. The exhausted heat from the FC is employed to heat water from city water network, which is then charged to a hot water storage tank. Hot water from this storage tank is utilized for fulfilling the dwelling hot water demand including air conditioner, floor heating, kitchen, bath, washing machine, toilet, etc. Any lack of hot water is compensated by a backup boiler inside the FC-CHP system (see Fig. \ref{EGWN}). 
The hot water amount charged to the storage tank 
is computed by
\begin{align}
\label{wt-2}
HT_{i,in}(t) = \frac{E_{i,fc}(t)\eta_{i,hr}(t)}{q_w(T_{i,ht}-T_{i,cn})}  
= \zeta_i \frac{\eta_{i,hr}(t)}{\eta_{i,e}(t)} P_{i,fc}(t)
\end{align}
where $\zeta_i \triangleq \frac{\Delta t \xi_e }{q_w(T_{i,ht}-T_{i,cn})\eta_{i,g2h}}$ is a positive constant. 
Because of the nonlinear dependence of the power and heat efficiencies $\eta_{i,e}(t)$ and $\eta_{i,hr}(t)$ to the FC output power $P_{i,fc}(t)$, $HT_{i,in}(t)$ is also a nonlinear function of $P_{i,fc}(t)$. Nevertheless, in the following we show that $HT_{i,in}(t)$ can be well approximated by a linear function of $P_{i,fc}(t)$.

\begin{figure}[htbp!]
		\centering
		\includegraphics[scale=0.4]{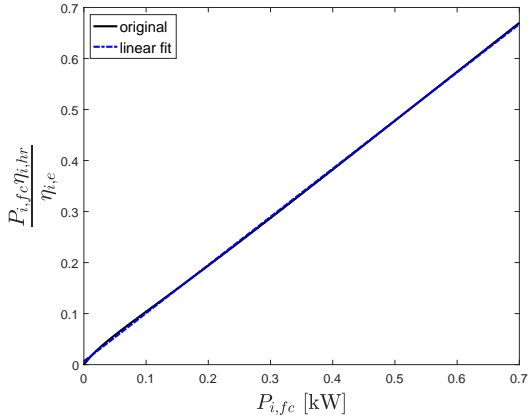}
		\caption{Linear approximation of hot water amount charged to storage tank.}
		\label{FC_heat_recovery_approx}
	\end{figure}

For a variety of FCs, the power and heat efficiencies $\eta_{i,e}(t)$ and $\eta_{i,hr}(t)$ have the exponential forms \cite{Tran18,Wakui10,Aki16} described as follows. 
\begin{equation}
	\label{FC-eff}
	\begin{aligned}
		\eta_{i,e}(t) &= a_{i,e} - b_{i,e}e^{-k_{i,e}\frac{P_{i,fc}(t)}{P_{i,fc}^{\max}}}, \\ a_{i,e} &= b_{i,e} + \eta_{i,e}^{0}, 
		 b_{i,e} = \frac{\eta_{i,e}^{\max}-\eta_{i,e}^{0}}{1-e^{-k_{i,e}}} \\
		\eta_{i,hr}(t) &= a_{i,hr} - b_{i,hr}e^{-k_{i,hr}\frac{P_{i,fc}(t)}{P_{i,fc}^{\max}}}, \\
		a_{i,hr} &= b_{i,hr} + \eta_{i,hr}^{0}, ~ b_{i,hr} = \frac{\eta_{i,hr}^{\max}-\eta_{i,hr}^{0}}{1-e^{-k_{i,hr}}}
	\end{aligned}
\end{equation}
where $P_{i,fc}^{\max}$ is the maximum rated output power of the FC in dwelling $i$; and $k_{i,e},\eta_{i,e}^{0},\eta_{i,e}^{\max},k_{i,hr},\eta_{i,hr}^{0},\eta_{i,hr}^{\max}$ are constant parameters. To linearize $HT_{i,in}(t)$, we approximate 
\begin{equation}
	\label{ht-eq1}
	\frac{\eta_{i,hr}(t) }{\eta_{i,e}(t)} P_{i,fc}(t) \approx \alpha_{i,wt}P_{i,fc}(t)+\beta_{i,wt}
\end{equation}
where $\alpha_{i,wt}$ and $\beta_{i,wt}$ are constants. As an example, using the data of a SOFC provided in \cite{Tran18}, the linear approximation above, conducted in MATLAB, gives us $\alpha_{i,wt}=0.9439$, $\beta_{i,wt}=0.006502$, with $95\%$ confidence bounds, which is depicted in Figure \ref{FC_heat_recovery_approx}. 

As seen in Figure \ref{FC_heat_recovery_approx}, this linear approximation is very close to the original nonlinear curve. Therefore, the hot water amount charged to the storage tank is mostly linearly proportional to the FC output power. This obviously will reduce significantly the complexity of the FC-CHP optimal energy management problem when the hot water trading \cite{Aki18,Aki16,Tran18} is considered, because the optimization problem is convexified and no efficiency matching algorithm is needed, unlike that in \cite{Tran18}. This point is further elaborated in Section \ref{ems-cvx}, where the nonlinear term $\frac{P_{i,fc}(t)}{\eta_{i,e}(t)}$ in the gas usage cost is also shown to be reasonably estimated as a linear function of the FC output power $P_{i,fc}(t)$.    

\begin{rem}
\label{rem1}
The linearization method shown in \eqref{ht-eq1} is general and applicable to other FCs. One may also use a piecewise linear approximation for $\frac{\eta_{i,hr}(t) }{\eta_{i,e}(t)} P_{i,fc}(t)$, nevertheless how many linear segments to be used would be case-sensitive. In addition, the FC output power $P_{i,fc}(t)$ needs to be measured to know which linear segment should be utilized, hence making the problem more complicated. Thus, in this research, we prefer the linearization of $\frac{\eta_{i,hr}(t) }{\eta_{i,e}(t)} P_{i,fc}(t)$ to the piecewise linear approximation. 

Similar remark applies for the linearization of the FC gas energy consumption shown later in Section \ref{ems-cvx}.
\end{rem}
Having the linear approximation of hot water storage -- FC output power relation, we can estimate quite precisely the storage tank level at each time step $t$. Accordingly, a smart FC-CHP controller, which memorizes electricity and water consumptions in previous days, will notify dwellings' owners once the hot water storage tanks are near their limits through in-dwelling monitors or applications in smart phones, tablets, etc., to release hot water in advance of users' desire, e.g. for bath tub, washing machines, etc. Such actions can also be automatically executed by the controller through pre-programmed functions. Thus, in this study we do not consider hot water in the FC-CHP optimal energy management, instead we focus on gas-electricity management with electricity trading between FC-CHP equipped dwellings.

\subsection{P2P Electricity Trading}

Considering the P2P energy trading between $n$ dwellings/peers during the time interval $[1,\bT]$. Denote $P_{ij}(t)$ the energy to be traded at time step $t$ between the $i$-th and $j$-th peers, where $P_{ij}(t)>0$ means peer $i$ buys electricity from peer $j$, and vice versa, $P_{ij}(t)<0$ means peer $i$ sells electricity to peer $j$. To simplify the trading of dwellings/peers, it is assumed that at each time step one dwelling/peer only buys or sells energy, but not to do both. For each dwelling/peer $i$, denote $\tN_i$ its neighboring set, i.e. the set of other peers it is communicated for energy trading. 
Denote $\tG$ the inter-peer communication graph. Due to the bilateral trading between peers, $\tG$ is undirected.  
Next, let $a_{ij}$ be elements of the adjacency matrix $\tA$, i.e. $a_{ij}=1$ if peers $i$ and $j$ are connected, and  $a_{ij}=0$ otherwise. 
The degree matrix $\tD$ is defined by $\tD=\di\{d_{i}\}_{i=1,\ldots,n}$, where $d_{i} \triangleq \sum_{j \in \tN_{i}}{a_{ij}}$. 
Then the Laplacian matrix $\mathcal{L}$ associated to $\mathcal{G}$ is defined by $\tL=\tD-\tA$.  

Let $n_{i} \triangleq |\tN_i|$, $P_i \in \bR^{n_{i}}$ be the vector of all $P_{ij}$ with $j \in \tN_i$, $P_{i,tr}(t)$ be dwelling $i$'s total traded power. Then $P_{i,tr}(t)=\mathbf{1}_{n_{i}}^TP_i(t)$.  
Next, denote $C_i(P_{i,tr}(t))$ the total cost of dwelling/peer $i$ for trading in the P2P market, which composes of the following two components. 
The first component is the utility function, 
\begin{equation}
\label{cost-1}
C_{i,1}(P_{i}(t)) = a_i(t) P_{i,tr}^2(t) + \tilde{b}_i(t) P_{i,tr}(t) + c_i(t) 
\end{equation}
The parameters $a_i(t), \tilde{b}_i(t), c_i(t)$ are only known for peer $i$, which are presented here as time-dependent parameters to reflect the time-varying and complex behaviors of dwellings/peers. 
The second element is the implementation cost for the traded powers to be physically executed through the power network, 
\begin{equation}
\label{cost-3}
C_{i,2}(P_{i}(t)) = \gamma P_{i,tr}(t)
\end{equation} 
where $\gamma>0$ is a fixed rate. 
Thus, summing up \eqref{cost-1} and \eqref{cost-3} gives us the following total cost of each peer in the P2P market 
\begin{equation}
\label{cost}
C_{i}(P_{i}(t)) = a_i(t) P_{i,tr}^2(t) + \hat{b}_i(t) P_{i,tr}(t) + c_i(t) 
\end{equation} 
where $\hat{b}_i(t) \triangleq \tilde{b}_i(t) + \gamma$. 

\begin{rem}
\label{rem2}
In the considering system, the hot water storage tank in fact is a thermal storage device which stores hot water output from the FC for later use. On the other hand, we do not consider electric storage devices, e.g. battery, due to the existence of the P2P electricity trading market between dwellings. Any lack or redundancy of electricity of any dwelling will be compensated through such P2P electricity trading market, or the grid. Therefore, electric storage devices are not needed to avoid increased investment cost for dwelling owners. 
\end{rem}

\subsection{System Constraints}

The first constraint is the power balance 
\begin{equation}
\label{e-1}
P_{i,fc}(t) + P_{i,tr}(t) + P_{i,grid}(t) = P_{i,dem}(t)  \;\forall \; t=1,\ldots,\bT
\end{equation}
The priority for compensating any lack of demand is from the P2P energy market. Grid power is bought only when powers from neighboring dwellings are not enough. 

Next, the electricity generated by a FC is bounded by
\begin{equation}
\label{e-fc}
P_{i,fc}^{\min}(t) \leq P_{i,fc}(t) \leq P_{i,fc}^{\max} \;\forall \; t=1,\ldots,\bT; \; i=1,\ldots,n
\end{equation}
Since FC is the main power source for dwellings, it is expected that the FC output power is enough for a dwelling power demand, i.e., 
\begin{equation}
\label{pmin-1}
P_{i,fc}^{\min}(t) \geq P_{i,dem}(t)
\end{equation}
However, this is not always true, because there is a possibility that a dwelling power demand could be greater than the FC maximum rated output power, i.e., $P_{i,dem}(t) \geq P_{i,fc}^{\max}$. In this case, we have 
\begin{equation}
\label{pmin-2}
P_{i,fc}^{\min}(t)= P_{i,fc}^{\max}
\end{equation}
The combination of \eqref{pmin-1} and \eqref{pmin-2} therefore gives us   
\begin{align}
\label{e-fc-bound}
P_{i,fc}^{\min}(t) \triangleq \min \{ P_{i,dem}(t), P_{i,fc}^{\max} \}
\end{align}
Lastly, the constraints on the maximum and minimum P2P electricity trading are determined by  
\begin{equation}
\label{e-2}
P_{i,tr}^{\min}(t) \leq P_{i,tr}(t) \leq P_{i,tr}^{\max}(t) \;\forall \; t=1,\ldots,\bT 
\end{equation}
where 
\begin{align}
\label{e-p2p-bounds}
P_{i,tr}^{\min}(t) &\triangleq \min \{ 0, P_{i,dem}(t) - P_{i,fc}^{\max} \}, \notag\\
 P_{i,tr}^{\max}(t) &\triangleq \max \{ 0, P_{i,dem}(t) - P_{i,fc}^{\max} \}
\end{align}

\subsection{Overall Optimization Problem}

Dwellings adjust the functionality of their FCs to minimize their total energy costs during the considered time period $[1,\bT]$, subject to the specified constraints, which results in the following optimization problem.
\begin{subequations}
\label{ems}
\begin{align}
\min ~ & \sum_{t=1}^{\bT} \sum_{i=1}^{n} C_{i,g}(t)+C_i(P_{i}(t)) \\
\textrm{s.t.} ~ & \eqref{e-fc}, \eqref{e-2} 
\end{align}
\end{subequations}
It is worth emphasizing again that electric demand in each dwelling is time-varying and inconsistent from one day to another. Moreover, electric demand prediction is often inexact. Therefore, solving \eqref{ems} for the whole time period $[1,\bT]$ at once is not a good choice in practice for intra-day energy management. Instead, it is better to solve\eqref{ems} at each time step $t$ so that the FC-CHPs in dwellings can be scheduled properly to the variation of electric demands. As such, the cooperative energy management problem \eqref{ems} should be solved one-time-step ahead. This means in the optimization problem to be solved for the next time step $t$, the values of all variables and costs up to the current time steps are known. Therefore, we only need to find the variables and costs at time step $t$, as shown in \eqref{ems-1}. 
\begin{subequations}
\label{ems-1}
\begin{align}
\min ~ & \sum_{i=1}^{n} C_{i,g}(t)+C_i(P_{i,tr}(t))    \\
\textrm{s.t.} 
~ & P_{i,fc}^{\min}(t) \leq P_{i,fc}(t) \leq P_{i,fc}^{\max} \\
~ & P_{i,tr}^{\min} \leq P_{i,tr}(t) \leq P_{i,tr}^{\max} \\
~ & P_{ij}(t) + P_{ji}(t) = 0 ~\forall \, j \in \tN_i, i,j=1,\ldots,n
\end{align}
\end{subequations} 
where $P_{i,fc}^{\min}(t)$, $P_{i,tr}^{\min}$, and $P_{i,tr}^{\max}$ are specified in \eqref{e-fc-bound} and \eqref{e-p2p-bounds}.

Note that \eqref{ems-1} is a non-convex problem due to the nonlinearity of the FC power efficiency $\eta_{i,e}(t)$ which causes the non-convexity of the cost function $C_{i,g}(t)$ in \eqref{g-cost}. Therefore, in this research we propose an approximation of $C_{i,g}(t)$ as a linear function of $P_{i,fc}(t)$, hence \eqref{ems-1} becomes convex.

\section{The Proposed Energy Management Approach}
\label{proposed-approach}

\subsection{Convexification of FC-CHP Optimal Energy Management Problem}
\label{ems-cvx}

\begin{figure}[htbp!]
		\centering
		\includegraphics[scale=0.4]{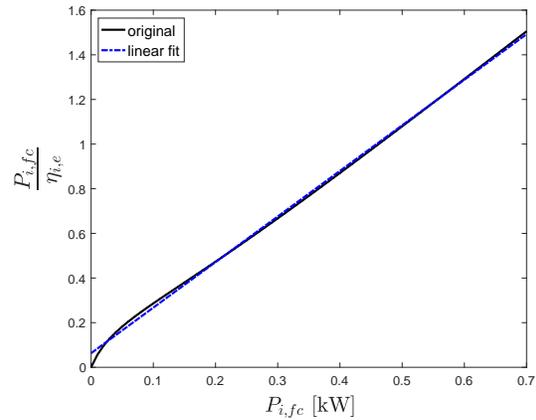}
		\caption{Linear approximation of the nonlinearity in FC gas energy consumption.}
		\label{FC_energy_approx}
	\end{figure}

As aforementioned, we aim at linearizing $C_{i,g}(t)$ to make \eqref{ems-1} a convex problem. To do so, $\frac{P_{i,fc}(t)}{\eta_{i,e}(t)}$ is linearized as follows,
\begin{align}
\label{g-lin}
\frac{P_{i,fc}(t)}{\eta_{i,e}(t)} \approx \alpha_{i,fc}P_{i,fc}(t)+\beta_{i,fc}
\end{align}
where $\alpha_{i,fc}$ and $\beta_{i,fc}$ are constants. 
Therefore, the FC-CHP gas energy consumption is approximated by 
\begin{align}
\label{g-2}
E_{i,fc}(t) \approx (\alpha_{i,fc}P_{i,fc}(t)+\beta_{i,fc}) \dfrac{\Delta t \xi_e}{\eta_{i,g2h}}
\end{align}
As an example, using the data of a SOFC provided in \cite{Tran18}, we first obtain the nonlinear curve for the ratio $\frac{P_{i,fc}(t)}{\eta_{i,e}}$ as shown by the solid black line in Figure \ref{FC_energy_approx}. Next, a linear fit conducted in MATLAB gives us the linear approximation depicted by the dash-dot blue line in Figure \ref{FC_energy_approx}, where $\alpha_{i,fc}=2.042$ and $\beta_{i,fc}=0.06323$, with $95\%$ confidence bounds.  

Note that the P2P trading between dwellings pushes their FCs to produce higher output powers so that they have powers to trade with the others, hence increases the FC power and heat efficiencies. Also, our proposed linear approximation is more accurate when the FC output power is higher (see Figures \ref{FC_heat_recovery_approx}--\ref{FC_energy_approx}), therefore it is suitable and reasonable for the system we are considering. Further, this approximation allows us to obtain a convexified problem without any efficiency matching algorithm (e.g. that in \cite{Tran18}), hence the complexity as well as the running time are much lower.

The convexified optimal energy management problem is described below, where the constant terms are removed from the objective function.
\begin{subequations}
\label{ems-2}
\begin{align}
\label{cost-t}
\min ~ & \sum_{i=1}^{n} \frac{\alpha_{i,fc}\Delta t \xi_e}{\eta_{i,g2h}}P_{i,fc} + a_i P_{i,tr}^2 + \hat{b}_i P_{i,tr}    \\
\label{fc-balance}
\textrm{s.t.} 
~ & P_{i,fc}^{\min} \leq P_{i,fc} \leq P_{i,fc}^{\max} \\
\label{trade-constraint-0}
~ & P_{i,tr}^{\min} \leq P_{i,tr} \leq P_{i,tr}^{\max} \\
~ & P_{ij} + P_{ji} = 0 ~\forall \, j \in \tN_i, i,j=1,\ldots,n
\end{align}
\end{subequations} 
We proceed by further elaborating on the optimization problem \eqref{ems-2} with two scenarios of dwelling demand. First, if $P_{i,dem} \geq P_{i,fc}^{\max}$, then the constraint \eqref{fc-balance} implies that $P_{i,fc}=P_{i,fc}^{\max}$, due to \eqref{e-p2p-bounds}, hence the first term on gas consumption cost in \eqref{cost-t} is constant. As a result, the local optimization problem at dwelling $i$ becomes a problem for only variables $P_{ij}$, i.e., for only P2P energy trading, as follows, where \eqref{e-p2p-bounds} is explicitly employed.
\begin{subequations}
\label{ems-3}
\begin{align}
\label{cost-t-1}
\min ~ &  a_i P_{i,tr}^2 + \hat{b}_i P_{i,tr}    \\
\label{trade-constraint-1}
\textrm{s.t.} ~ & 0 \leq  P_{i,tr} \leq P_{i,dem} - P_{i,fc}^{\max}  \\
~ & P_{ij} + P_{ji} = 0 ~\forall \, j \in \tN_i, i,j=1,\ldots,n
\end{align}
\end{subequations} 
Second, if $P_{i,dem} < P_{i,fc}^{\max}$, then dwelling $i$ becomes a potential seller in the P2P energy market with $P_{i,tr}=P_{i,dem}-P_{i,fc}$. Then the local optimization problem at dwelling $i$ again is an optimization problem of only variables $P_{ij}$ as in the following, where $P_{i,fc}$ in \eqref{ems-2} is substituted by $P_{i,dem}-P_{i,tr}$, and \eqref{e-p2p-bounds} is explicitly utilized.
\begin{subequations}
\label{ems-4}
\begin{align}
\label{cost-t-2}
\min ~ & a_i P_{i,tr}^2 + \left(\hat{b}_i-\frac{\alpha_{i,fc}\Delta t \xi_e}{\eta_{i,g2h}}\right) P_{i,tr}    \\
\label{trade-constraint-2}
\textrm{s.t.} ~  &  P_{i,dem}(t) - P_{i,fc}^{\max} \leq P_{i,tr} \leq 0  \\
~ & P_{ij} + P_{ji} = 0 ~\forall \, j \in \tN_i, i,j=1,\ldots,n
\end{align}
\end{subequations}  
Thus, our considering optimal energy management for a group of dwellings equipped with FC-CHPs is in fact reduced to an optimal P2P energy trading which is a quadratic convex problem having the following form.
\begin{subequations}
\label{ems-5}
\begin{align}
\label{cost-p2p}
\min ~ &  \sum_{i=1}^{n} a_i P_{i,tr}^2 + b_i P_{i,tr}    \\
\label{trade-constraint}
\textrm{s.t.} ~ & P_{i,tr}^{\min} \leq P_{i,tr} \leq P_{i,tr}^{\max} \\
~ & P_{ij} + P_{ji} = 0 ~\forall \, j \in \tN_i, i,j=1,\ldots,n
\end{align}
\end{subequations} 
where $b_i$, $P_{i,tr}^{\min}$, and $P_{i,tr}^{\max}$ are determined in \eqref{ems-3} or \eqref{ems-4}.   
A distributed P2P energy trading mechanism will be proposed in the next section to solve \eqref{ems-5}. 

\subsection{Distributed P2P Mechanism}

In the following, a distributed and parallel ADMM approach is proposed to solve the mathematical programming \eqref{ems-5}. The advantage of this approach is that it allows each peer/agent to solve its own local optimization problem while negotiating with other peers/agents to eventually reach the solution of the global optimization problem \eqref{ems-5}. Thus, the communication and computation burden at a centralized entity is avoided, and the privacy of each peer/agent can be guaranteed. 

Denote $m \triangleq \sum_{i=1}^{n} n_{i}$ and $P \in \bR^{m}$ the vector of all $P_i,i=1,\ldots,n$. Then let us define the sets of coupling and local constraints as in \eqref{coupling-set} and \eqref{local-set}, respectively. 
\begin{align}
\label{coupling-set}
\Omega_{g} \triangleq & \left \{ P \in \bR^{m}: P_{ij}(t) + P_{ji}(t) = 0 ~\forall \, j  \in \tN_i \right \} 
\end{align}
\begin{align}
\label{local-set}
\Omega_{\ell} \triangleq & \left \{ P \in \bR^{m}: P_{i,tr}^{\min} \leq \mathbf{1}_{n_i}^T P_i \leq P_{i,tr}^{\max}  \right \}
\end{align}
For those sets, the following indicator functions are defined. 
\begin{align}
\label{I-eq}
I_{g}(P) \triangleq \left \{ 
\begin{array}{rl}
	0: & P \in \Omega_{g} \\
	+\infty: & P \notin  \Omega_{g}
\end{array}
\right., ~
I_{\ell}(P) \triangleq \left \{ 
\begin{array}{rl}
	0: & P \in \Omega_{\ell} \\
	+\infty: & P \notin  \Omega_{\ell}
\end{array}
\right.
\end{align}
Now, by utilizing a new variable $X \in \bR^{m}$, the optimization problem \eqref{ems-1} is rewritten such that equality and inequality constraints are separated into different sets corresponding to different variables $P$ and $X$, as follows. 
\begin{subequations}
\label{p2p-1}
\begin{align}
\min \, & \sum_{i=1}^{n}  C_i(P_i) + I_{g}(P) + I_{\ell}(X) \\
\textrm{s.t.} \; & P - X = 0 \\
\; & P \in \Omega_{g}, ~ X \in \Omega_{\ell}
\end{align}
\end{subequations}  
Obviously, \eqref{p2p-1} is in the standard form of the ADMM method \cite{Boyd:2011}, which solves \eqref{p2p-1} iteratively. Nevertheless, the classical ADMM method in \cite{Boyd:2011} is centralized and the updates of variables are in order. Therefore, in this research, we propose a novel ADMM approach that solves \eqref{p2p-1} in a fully distributed manner, and variables are updated in parallel.  
Define the following augmented Lagrangian, 
\begin{align*}
		L_{\rho}(P,X,u) = & \sum_{i=1}^{n} C_i(P_i) + I_{g}(P) + I_{\ell}(X) + \frac{\rho}{2}\|P-X+u\|_{2}^{2}, 		
\end{align*}
where $\rho>0$ is a scalar penalty parameter and $u \in \mathbb{R}^{m}$ is called the scaled Lagrange (dual) multiplier \cite{Boyd:2011}. 
Next, the variables $P,X,u$ are 
computed at each algorithm iteration $k+1$ by solving the following sub-problems,
	\begin{align}
		\label{var-update}
		X^{k+1} &\triangleq \argmin_{X \in \Omega_{\ell}}\left[L_{\rho}(P^{k},X,u^{k}) \right. \notag \\
		& \quad \left. + \frac{1}{2}(X-X^{k})^{T}\Psi(X-X^{k})\right] \notag \\	
		P^{k+1} &\triangleq \argmin_{P \in \Omega_{g}}\left[L_{\rho}(P,X^{k},u^{k}) \right. \notag \\
		& \quad \left. + \frac{1}{2}(P-P^{k})^{T}\Phi(P-P^{k})\right] \notag \\
		u^{k+1} &\triangleq u^{k} - \kappa \rho(P^{k}-X^{k})  
	\end{align}		
in which $\Phi,\Psi,\kappa>0$ satisfy
\begin{equation}
\label{prox-cond}
	\Phi \succ \rho(\frac{1}{\mu_1}-1)I, ~\Psi \succ \rho(\frac{1}{\mu_2}-1)I, ~\mu_1+\mu_2 < 2-\kappa
\end{equation}
for some $\mu_1>0,\mu_2>0$. Condition \eqref{prox-cond} was proved to be sufficient for the convergence of the above variables update \cite{Deng2017}. Note that the selection of $\Phi,\Psi,\kappa$ to fulfill \eqref{prox-cond} is not unique. One simple way is to let $\Phi=\phi I$, $\Psi=\psi I$ such that 
\begin{equation}
\label{prox-cond-1}
	\phi > \rho(\frac{1}{\mu_1}-1), ~\psi > \rho(\frac{1}{\mu_2}-1), 
	~\mu_1+\mu_2 < 2-\kappa
\end{equation} 
Unlike the ordered updates in the classical centralized ADMM \cite{Boyd:2011} and the existing distributed ADMM methods (e.g. \cite{Nguyen-TSG17}), each variable in (\ref{var-update}) at each iteration is completely independent of the other two variables at the same iteration. Therefore, problem variables are computed in parallel at every agent. Moreover, our proposed ADMM approach advances that in \cite{Deng2017}, which also updates variables in parallel, by allowing sparse connection between peers/agents, while that in \cite{Deng2017} requires all-to-all inter-agent connection. 

The updates of variables $P$ and $X$ in \eqref{var-update} are explicitly presented in the next sections.

\subsection{The Update for Variable $X$}

The update for $X^{k+1}$ in \eqref{var-update} is derived by solving the following optimization problem.
\begin{subequations}
\label{p2p-X}
\begin{align}
\min \, & \frac{\rho}{2}\|P^k-X+u^k\|_{2}^{2} + \frac{\psi}{2}\|X-X^{k}\|_2^2 \\
\textrm{s.t.} \; & P_{i,tr}^{\min} \leq \mathbf{1}_{n_{i}}^T X_i \leq P_{i,tr}^{\max} ~\forall \, i=1,\ldots,n
\end{align}
\end{subequations} 
Note that there is one more constraint on the positiveness or negativeness of each $X_i,i=1,\ldots,n$, depending on whether the $i$-th peer/agent at the next time slot will perform as a buyer or seller. In any case, \eqref{p2p-X} is a quadratic convex problem and is decompsable to each peer/agent, i.e. it is fully decentralized, hence it can be easily solved by any off-the-self software embedded in each peer/agent, e.g., CVX \cite{CVX}. 

\subsection{The Update for Variable $P$}

To obtain the update for $P^{k+1}$ in \eqref{var-update}, we need to solve the following mathematical programming.
\begin{subequations}
\label{p2p-P}
\begin{align}
\min \, & \sum_{i=1}^{n} C_i(P_i) + \frac{\rho}{2}\|P-X^k+u^k\|_{2}^{2} + \frac{\phi}{2}\|P-P^{k}\|_2^2 \\
\label{p2p-balance}
\textrm{s.t.} \; & P_{ij} + P_{ji} = 0 ~\forall \, i=1,\ldots,n; \ j \in \tN_i
\end{align}
\end{subequations}
Denote $\lambda_{ij}>0$ the Lagrange multiplier associated with the constraint \eqref{p2p-balance}, and $\lambda_i \in \bR^{n_{i}}$ the vector of all $\lambda_{ij}$ with $j \in \tN_i$. Since \eqref{p2p-P} is a convex optimization problem with quadratic cost function and linear equality, the strong duality holds and KKT 
conditions apply. Therefore, we obtain from \eqref{p2p-P} that
\small
\begin{align}
\label{price-eq}
\lambda_{ij}^{k+1}  =& \; \frac{\partial}{\partial P_{ij}} \left(\sum_{i=1}^{n} C_i(P_i) + \frac{\rho}{2}\|P-X^k+u^k\|_{2}^{2} \right. \notag \\
& \qquad \left. \left. + \frac{\phi}{2}\|P-P^{k}\|_2^2 \right) \right|_{P_{ij}=P_{ij}^{k+1}}  \notag \\
 =& \; 2a_iP_{i,tr}^{k+1} + (\rho+\phi)P_{ij}^{k+1} + v_{ij}^{k}
\end{align}
\normalsize
where $v_{ij}^{k} \triangleq  b_i + d_{ij} + \rho(-X_{ij}^k+u_{ij}^k) - \phi P_{ij}^k$. 
Next, due to the bilateral trading constraint \eqref{p2p-balance}, the Lagrange multipliers and the traded powers must satisfy the following constraints.
\begin{equation}
\label{price-constraint}
\lambda_{ij}^{k+1} = \lambda_{ji}^{k+1}, P_{ij}^{k+1} = -P_{ji}^{k+1} ~\forall \; j \in \tN_i
\end{equation} 
Here, $\lambda_{ij}^{k+1}$ is considered to be the trading price between the $i$-th and $j$-th peers. 
Interestingly, equation \eqref{price-eq} reveals the relation between individually traded price between a pair of peers/agents with its associated traded power and the total available powers of those peers/agents. 

Let us denote
\begin{align*}
\Gamma &\triangleq (\rho+\phi)\mathrm{diag}\{\frac{1}{a_i}\}_{i=1,\ldots,n}, ~ \tilde{P}_{i,tr}^{k+1} \triangleq  2a_iP_{i,tr}^{k+1},  \\
\hat{v}_i^{k+1} &\triangleq \sum_{j \in \tN_i} v_{ji}^{k+1}, ~ \hat{v}^{k+1} \triangleq \begin{bmatrix} \hat{v}_1^{k+1},\cdots,\hat{v}_n^{k+1} \end{bmatrix}^T,\\ 
\tilde{v}_i^{k+1} &\triangleq \sum_{j \in \tN_i} v_{ij}^{k+1}, ~ 
 \tilde{v}^{k+1} \triangleq \begin{bmatrix} \tilde{v}_1^{k+1},\cdots,\tilde{v}_n^{k+1} \end{bmatrix}^T 
\end{align*}
Utilizing \eqref{price-constraint} and \eqref{price-eq}, we can easily obtain 
\begin{subequations}
\begin{align}
\label{opt-var-1}
P_{ij}^{k+1} &= \frac{v_{ji}^{k+1}+\tilde{P}_{j,tr}^{k+1}-v_{ij}^{k+1}-\tilde{P}_{i,tr}^{k+1}}{2(\rho+\phi)} \\
\label{opt-var-11}
\lambda_{ij}^{k+1} &= \frac{v_{ji}^{k+1}+\tilde{P}_{j,tr}^{k+1}+v_{ij}^{k+1}+\tilde{P}_{i,tr}^{k+1}}{2} 
\end{align}
\end{subequations}
The convergence of the proposed ADMM algorithm follows that provided in \cite{Deng2017}, hence we omit the proof here for brevity. Next, 
substituting the optimal solutions to \eqref{price-eq} leads to $\lambda_{ij}^{\ast}  = 2a_iP_{i,tr}^{\ast} + b_i + d_{ij} + \rho u_{ij}^{\ast}$. 
Then summing up \eqref{opt-var-1} for all $j \in \tN_i$ leads to
\small
\begin{align*}
P_{i,tr}^{k+1}  
= \frac{\displaystyle \sum_{j \in \tN_i} v_{ji}^{k+1} + \sum_{j \in \tN_i} \tilde{P}_{j,tr}^{k+1} - \sum_{j \in \tN_i} v_{ij}^{k+1} - n_{i}\tilde{P}_{i,tr}^{k+1}}{2(\rho+\phi)} 
\end{align*}
\normalsize
which is equivalent to
\begin{align}
\label{opt-var-2}
2(\rho+\phi+n_{i}a_i)P_{i,tr}^{k+1} - \sum_{j \in \tN_i} 2a_jP_{j,tr}^{k+1} 
= \sum_{j \in \tN_i} v_{ji}^{k+1}  - \sum_{j \in \tN_i} v_{ij}^{k+1}  
\end{align}
Stacking \eqref{opt-var-2} with $i=1,\ldots,n$ results in 
\begin{equation}
\label{opt-P}
(\tL + \Gamma) \tilde{P}_{tr}^{k+1} = \hat{v}^{k+1} - \tilde{v}^{k+1}
\end{equation}
Equations \eqref{opt-P}, \eqref{opt-var-1}, and \eqref{opt-var-11} give us the updates for variables $P_{ij}^{k+1}$ and the P2P energy prices $\lambda_{ij}^{k+1}$. 

\begin{table*}[htpb!]	
	\caption{Time-varying structure of P2P electricity trading between 6 houses.}
	\begin{center}
		\scalebox{0.8}{
		\begin{tabular}{|c|c|c|c|c|}
			\hline
			Time steps & 2, 16, 32, 33, 36, 37, 38, 47 & 6, 27, 28 & 7 & 8  \\
			\hline
			Sellers & 2, 3, 4, 5, 6 & 1, 2, 4, 5, 6 & 1, 4, 5, 6  & 1, 2, 5 \\
			\hline			
			Buyers & 1 & 3 & 2, 3 & 3, 4, 6 \\
			\hline \hline			
			Time steps & 9, 10 & 11 & 12 & 13 \\
			\hline
			Sellers & 1, 2, 4, 5 & 1, 2 & 2 & 2, 3, 5  \\
			\hline			
			Buyers & 3, 6 & 3, 4, 5, 6 & 1, 3, 4, 5, 6 & 1, 4, 6 \\
			\hline\hline
			Time steps & 14, 15 & 17 & 18, 19, 20, 21 & 29 \\
			\hline
			Sellers & 2, 3 & 2, 3, 5, 6 & 1, 2, 3, 5, 6 & 2, 4, 5, 6  \\
			\hline			
			Buyers & 1, 4, 5, 6 & 1, 4 & 4 & 1, 3 \\
			\hline \hline						
			Time steps & 42, 43 & 44 & 45 & 46 \\
			\hline
			Sellers & 1, 3, 4, 5, 6 & 1, 3, 4, 6 & 1, 2, 3, 4, 6 & 2, 3, 4, 6  \\
			\hline			
			Buyers & 2 & 2, 5 & 5 & 1, 5 \\
			\hline 							
		\end{tabular}
		}
	\end{center}	
	\label{p2p-structure}
\end{table*}

\section{Case Studies}
\label{cases}

\begin{figure}[htbp!]
		\centering
		\includegraphics[scale=0.45]{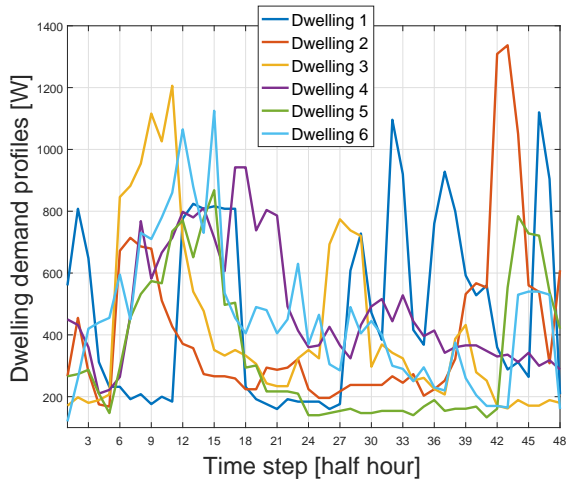}
		\caption{Electric consumption patterns of 6 houses.}
		\label{homes_dem_original}
	\end{figure}

In this section, we consider a group of 6 houses, each house is equipped with one SOFC-CHP working in the range [50W, 700W], and the gas-to-hydrogen processing efficiency $\eta_{i,g2h}=95\%$. 
One-day electric consumption data of these 6 houses with half-hour resolution are adapted from a Toyota project.  
Due to distinct consumption patterns of different houses (see Figure \ref{homes_dem_original}), P2P electricity trading between them is obviously possible. Additionally, the role of each house as a buyer or seller is time-varying, which may change from one time step to another, depending on its predicted electricity demand at the next time step. More specifically, if the predicted power demand of a dwelling is greater than 700W, then it will be a buyer. On the other hand,  if the predicted power demand of a dwelling is smaller than 700W, then it will be a seller. As such, the interconnection structure in the P2P energy market is also time-varying (see Table \ref{p2p-structure}). 
This interesting characteristic of P2P energy markets is very different from conventional energy markets where suppliers and consumers are fixed, and hence is worth investigating. 

	\begin{figure}[htbp!]
		\centering
		\includegraphics[scale=0.36]{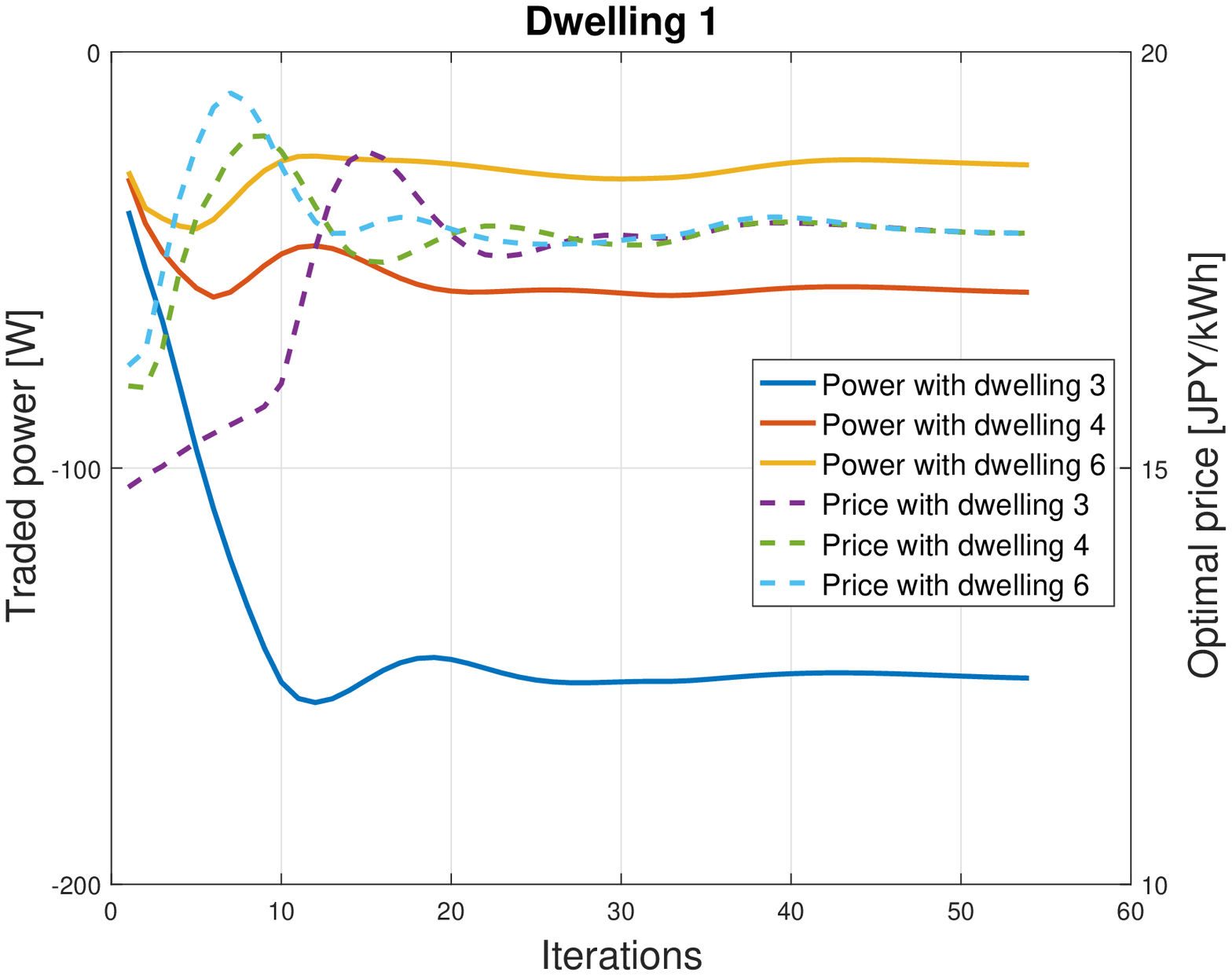}
	%
		\centering
		\includegraphics[scale=0.36]{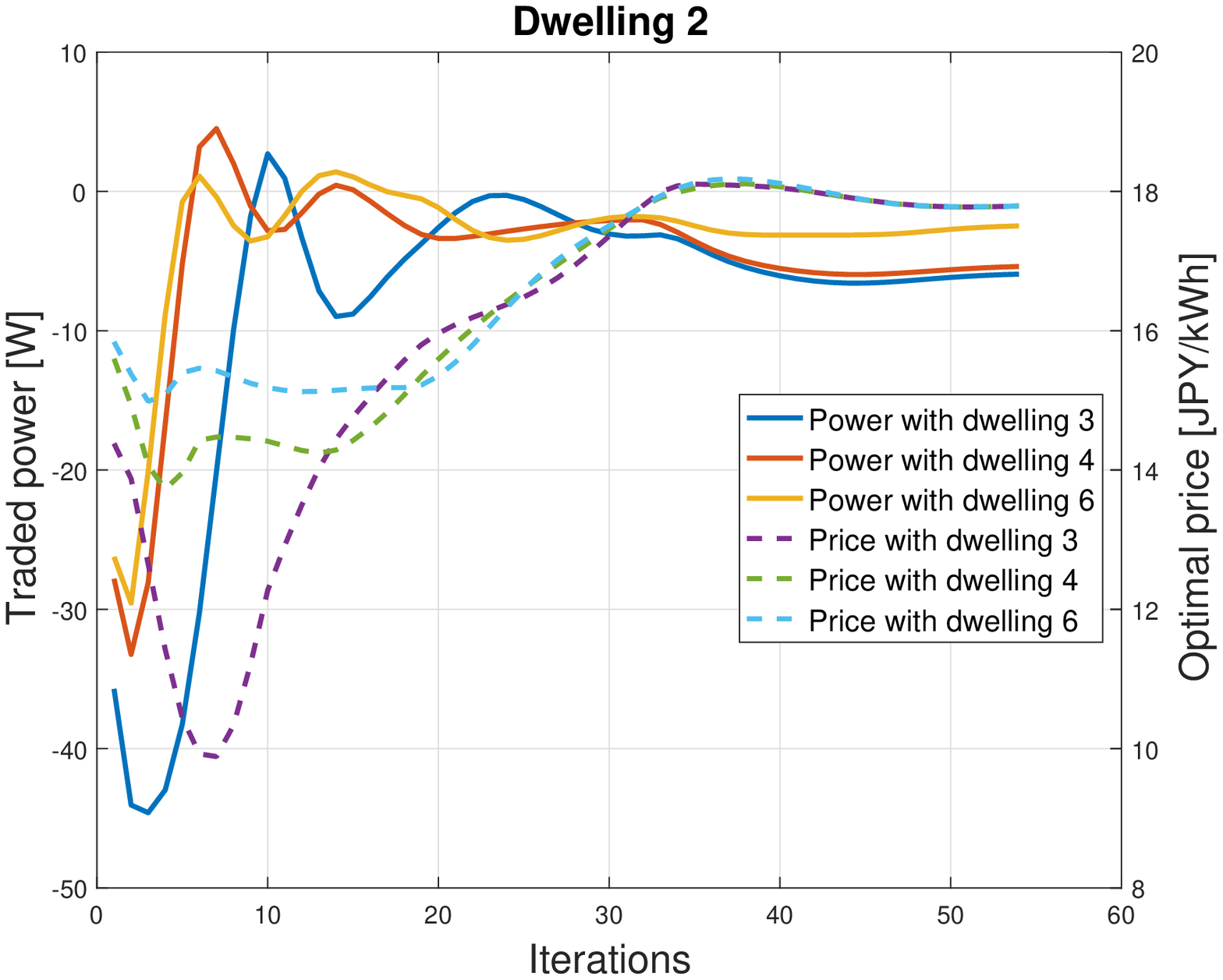}
		\caption{Power trading of house 1 (left) and house 2 (right) by the proposed distributed P2P mechanism.}
		\label{house2_case1}
	\end{figure}

Similarly to rooftop solar power generation, FC-CHPs should be subsidized to increase their deployment for reducing carbon emission. Suppose that the subsidized unit gas cost to produce 1kWh at the FC minimum and maximum power efficiency is 11.89 JPY and 20.31 JPY. Note that the electricity rate for households from the grid in Japan in in ladder-form starting around 20 JPY/kWh. 
Thus, dwellings can set their utility function parameters $a_i$ and $b_i$ such that the optimal P2P energy price is between those unit gas and electricity prices. How to select $a_i$ and $b_i$ is out of scope of this paper and will be presented in another work.

	\begin{figure}[htbp!]
		\centering
		\includegraphics[scale=0.36]{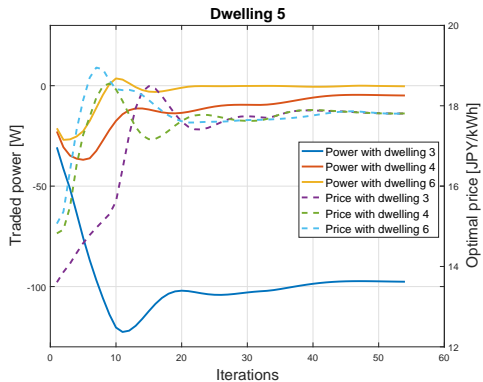}
		\caption{Power trading of house 5 by the proposed distributed P2P mechanism.}
		\label{house5_case1}
	\end{figure}

Next, employing our proposed optimal energy management strategy with P2P energy trading, the simulation results for time slot 8 are shown in Figures \ref{house2_case1}--\ref{house5_case1}, which requires highest number of iterations to converge. At this time slot, houses 1, 2, and 5 are potential sellers (c.f. Table \ref{p2p-structure}) with capacities 492W, 14W, and 168W, respectively, and house 3, 4, and 6 are buyers (c.f. Table \ref{p2p-structure}) with 254W, 68W, and 30W over their rated FC powers, respectively.


As observed from Figures \ref{house2_case1}--\ref{house5_case1}, energy is successfully traded between each of houses 1, 2 and the three buying houses, but house 5 only sells energy to house 3 and house 4. 
The optimal energy price is 17.8 JPY/kWh.  
The total sold power of houses 1, 2, and 5 are 235W, 14W, and 103W, respectively, whereas the total bought power of houses 3, 4, and 6 are 254W, 68W, and 30W, respectively. Thus, through the P2P energy market, houses 3, 4, and 6 can buy enough powers they need from houses 1, 2, and 5, and no power from the grid is needed. 

Computational times in this scenario, conducted on a computer having Intel Core i7-6700K CPU 4GHz and 64GB RAM, without counting communication times between the central unit and peers/agents (in centralized method), or between peers/agents (in distributed method), are as follows. 
\begin{itemize}
	\item Centralized ADMM method, i.e., by solving \eqref{ems-5} at a central unit: 13.42 s.
	\item Distributed ADMM approach with ordered update of variables in \cite{Nguyen-TSG17}: 11.08 s for each peer/agent.
	\item Distributed ADMM approach with parallel update of variables proposed in the current work: 10.67 s for each peer/agent.
\end{itemize}
These results clearly show the advantage of our proposed approach on saving computational effort, especially when the number of peers/agents is high. 

\begin{figure}[htbp!]
		\centering
		\includegraphics[scale=0.42]{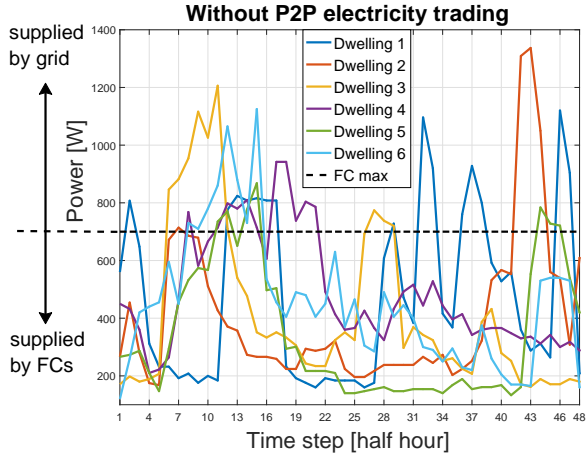}
		\caption{Electricity supplies for 6 houses if no P2P electricity market exists.}
		\label{homes_dem}
	\end{figure}

\begin{figure}[htbp!]
		\centering
		\includegraphics[scale=0.42]{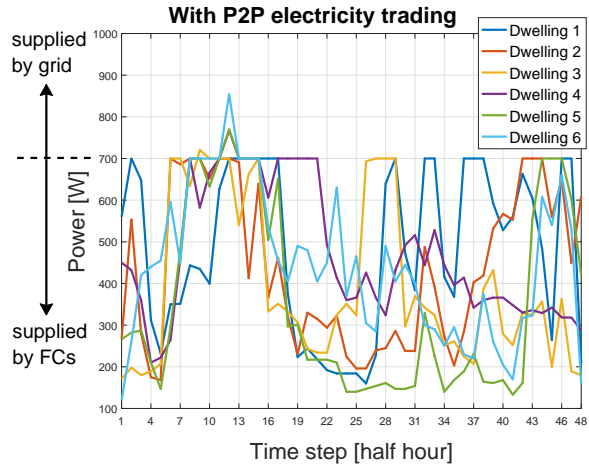}
		\caption{Electricity supplies for 6 houses if there is a P2P electricity market.}
		\label{homes_dem_new}
	\end{figure}

Note that the trading above occurs at time slot 8, i.e. at 3:30 am, hence this P2P energy trading can be made autonomously in the home energy management system without human intervention at any time during the day. Therefore, simulation results can be obtained at other time slots similarly, and due to the space limitation, we do not display all of their details here.   
Instead, we solve 31 optimization problems associated to 31 time steps with different P2P structures, at which electric demand of one or several houses exceeds the FC rated power, as shown in Table \ref{p2p-structure}. 
Then the electricity supplies from the dwelling FCs and from the grid are depicted in Figure \ref{homes_dem} and Figure \ref{homes_dem_new}, when a P2P electricity market exists and does not exist, respectively, which reveal significant differences.

When there is no P2P electricity market between dwellings, Figure \ref{homes_dem} shows that several dwellings need to buy much power from the grid, whereas the other dwelling FCs work at low outputs, e.g. at time steps 7--11, 42--47. On the other hand, in presence of a P2P electricity market between dwellings, it can be clearly seen from Figure \ref{homes_dem_new} that very few power amounts are bought from the grid by the dwellings. For most of the time steps, the electricity demand of dwellings can be fulfilled by their own FCs or by buying from other dwelling FCs through the P2P electricity market. Only at the time step 12, grid power is bought by dwellings 6 and 5, because the total selling power from the FCs in dwellings 1--4 are not enough. 

Further, when a P2P electricity market exists, FC units work at as high output power as possible in order to have power for selling, as seen in Figure \ref{homes_dem_new}. Thus, the linearization of FC hot water output and FC gas energy consumption presented in Section \ref{fc} and Section \ref{ems-cvx} are reasonable. In addition, P2P electricity market facilitates the deployment of DERs including FC-CHP and local energy consumption, hence reduces problems caused by reverse power flows to the bulk grid as well as decreases the power withdrawn from the bulk grid, especially during peak periods.

	\begin{figure}[htbp!]
		\centering
		\includegraphics[scale=0.4]{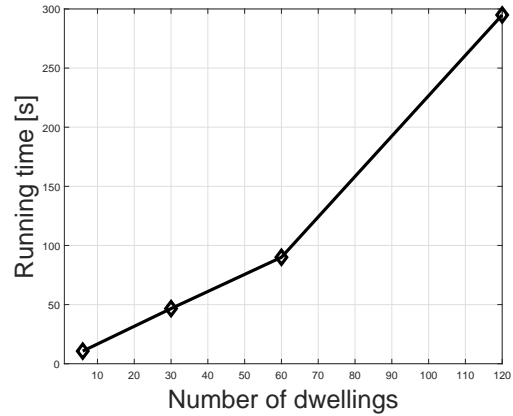}
		\caption{Illustration for the scalability of the proposed approach.}
		\label{admm_runtime_fc}
	\end{figure}
	
Lastly, to investigate how well the proposed approach works as the number of dwelling rises, the considered dwellings are duplicated to increase system size. Computational time for each peer/agent versus system size are then provided in Figure \ref{admm_runtime_fc}, which confirm the scalability of the proposed approach.

\section{Conclusion}
\label{sum}

This paper proposes a novel approach for optimal energy management in a local group of residential demand units equipped with FC-CHP systems and a P2P energy trading platform. By employing a linearization method for the FC hot water output and the FC gas energy consumption, the non-convex nonlinear optimization problem arising from such optimal energy management is convexified, hence much computational time and effort can be saved. 
Then an ADMM-based distributed and parallel method is proposed to solve the convexified optimization problem at each house. A case study using one-day realistic electricity demand from six houses with 30-minute sampling illustrates the effectiveness of the proposed approach on solving the optimal P2P energy trading problem and its positive impacts on the FCs operation and houses' electricity demands. The reasonably fast convergence and scalability of the proposed ADMM approach are also demonstrated.

In the future research, other DERs such as rooftop solar with on-site battery storage, heat pump water heater, or electric vehicles, will also be investigated to take into account the diversification of self-generation and self-consumption systems at local communities. Additionally, local heat demand in each dwelling will also be studied together with the P2P electricity trading to explicitly account for other heat consumption rather than hot water.

\section*{Acknowledgement}
The first author's research is financially supported by JSPS Kakenhi Grant Number 19K15013.


\bibliographystyle{plain}
\bibliography{dp}

\begin{thebibliography}{10}

\bibitem{CVX}
{\em {CVX: Matlab software for disciplined convex programming}}.

\bibitem{Aki16}
H.~Aki, T.~Wakui, and R.~Yokoyama.
\newblock {Development of an energy management system for optimal operation of
  fuel cell based residential energy systems}.
\newblock {\em {International Journal of Hydrogen Energy}}, 41:20314--20325,
  2016.

\bibitem{Aki18}
H.~Aki, T.~Wakui, R.~Yokoyama, and K.~Sawada.
\newblock {Optimal management of multiple heat sources in a residential area by
  an energy management system}.
\newblock {\em {Energy}}, 153:1048--1060, 2018.

\bibitem{Arsalis19}
A.~Arsalis.
\newblock {A comprehensive review of fuel cell-based
  micro-combined-heat-and-power systems}.
\newblock {\em {Renewable and Sustainable Energy Reviews}}, 105:391--414, 2019.

\bibitem{Baez-Gonzalez18}
P.~Baez-Gonzalez, E.~Rodriguez-Diaz, J.~C. Vasquez, and J.~M. Guerrero.
\newblock {Peer-to-Peer Energy Market for Community Microgrids}.
\newblock {\em {IEEE Electrification Magazine}}, 6(4):102--107, 2018.

\bibitem{Baroche19}
T.~Baroche, P.~Pinson, R.~L.G. Latimier, and H.~B. Ahmed.
\newblock {Exogenous Cost Allocation in Peer-to-Peer Electricity Markets}.
\newblock {\em {IEEE Transactions on Power Systems}}, 34(4):2553 -- 2564, 2019.

\bibitem{Boyd:2011}
S.~Boyd, N.~Parikh, E.~Chu, B.~Peleato, and J.~Eckstein.
\newblock Distributed optimization and statistical learning via the alternating
  direction method of multipliers.
\newblock {\em Foundations and Trends in Machine Learning}, 3(1):1--122, 2011.

\bibitem{Cadre20}
H.~Le Cadre, P.~Jacquot, C.~Wan, and C.~Alasseur.
\newblock {Peer-to-peer electricity market analysis: From variational to
  Generalized Nash Equilibrium}.
\newblock {\em {European Journal of Operational Research}}, 282:753--771, 2020.

\bibitem{SCui20}
S.~Cui, Y-W. Wang, Y.~Shi, and J-W. Xiao.
\newblock {A New and Fair Peer-to-Peer Energy Sharing Framework for Energy
  Buildings}.
\newblock {\em {IEEE Transactions on Smart Grid}}.

\bibitem{SCui19}
S.~Cui, Y-W. Wang, and J-W. Xiao.
\newblock {Peer-to-Peer Energy Sharing Among Smart Energy Buildings by
  Distributed Transaction}.
\newblock {\em {IEEE Transactions on Smart Grid}}, 10(6):6491--6501, 2019.

\bibitem{Deng2017}
W.~Deng, M-J. Lai, Z.~Peng, and W.~Yin.
\newblock Parallel multi-block admm with $o(1/ k)$ convergence.
\newblock {\em Journal of Scientific Computing}, 71(2):712--736, 2017.

\bibitem{DOE-CHP}
{Department of Energy, USA}.

\bibitem{Ellamla15}
H.~R. Ellamla, I.~Staffell, P.~Bujlo, B.~G. Pollet, and S.~Pasupathi.
\newblock {Current status of fuel cell based combined heat and power systems
  for residential sector}.
\newblock {\em {Journal of Power Sources}}, 293:312--328, 2015.

\bibitem{GuerreroA19}
J.~Guerrero, A.~C. Chapman, and G.~Verbic.
\newblock {Decentralized P2P Energy Trading under Network Constraints in a
  Low-Voltage Network}.
\newblock {\em {IEEE Transactions on Smart Grid}}, 2019.
\newblock {(Accepted). DOI: 10.1109/TSG.2018.2878445}.

\bibitem{JLPGA}
{Japan LP Gas Association}.

\bibitem{NLiu18}
N.~Liu, X.~Yu, W.~Fan, C.~Hu, T.~Rui, Q.~Chen, and J.~Zhang.
\newblock {Online Energy Sharing for Nanogrid Clusters: A Lyapunov Optimization
  Approach}.
\newblock {\em {IEEE Transactions on Smart Grid}}, 9(5):34624--4636, 2018.

\bibitem{NLiu17}
N.~Liu, X.~Yu, C.~Wang, C.~Li, L.~Ma, and J.~Lei.
\newblock {Energy-Sharing Model With Price-Based Demand Response for Microgrids
  of Peer-to-Peer Prosumers}.
\newblock {\em {IEEE Transactions on Power Systems}}, 32(5):3569--3583, 2017.

\bibitem{Khorasany19}
{M. Khorasany and Y. Mishra and G. Ledwich}.
\newblock {A Decentralised Bilateral Energy Trading System for Peer-to-Peer
  Electricity Markets}.
\newblock {\em {IEEE Transactions on Industrial Electronics}},
  67(6):4646--4657, 2019.

\bibitem{LMa16}
L.~Ma, N.~Liu, J.~Zhang, W.~Tushar, and C.~Yuen.
\newblock {Energy Management for Joint Operation of CHP and PV Prosumers Inside
  a Grid-Connected Microgrid: A Game Theoretic Approach}.
\newblock {\em {IEEE Transactions on Industrial Informatics}},
  12(5):1930--1942, 2016.

\bibitem{Moret19}
F.~Moret and P.~Pinson.
\newblock {Energy Collectives: a Community and Fairness based Approach to
  Future Electricity Markets}.
\newblock {\em {IEEE Transactions on Power Systems}}, 2019.
\newblock {(Accepted). DOI: 10.1109/TPWRS.2018.2808961}.

\bibitem{MorstynP2P18}
T.~Morstyn, N.~Farrell, S.~J. Darby, and M.~D. McCulloch.
\newblock {Using peer-to-peer energy-trading platforms to incentivize prosumers
  to form federated power plants}.
\newblock {\em {Nature Energy}}, 3:94--101, 2018.

\bibitem{MorstynP2P19a}
T.~Morstyn and M.~D. McCulloch.
\newblock {Multi-Class Energy Management for Peer-to-Peer Energy Trading Driven
  by Prosumer Preferences}.
\newblock {\em {IEEE Transactions on Power Systems}}, 2019.
\newblock {(Accepted). DOI: 10.1109/TPWRS.2018.28344721}.

\bibitem{MorstynP2P19b}
T.~Morstyn, A.~Teytelboym, and M.~D. McCulloch.
\newblock {Bilateral Contract Networks for Peer-to-Peer Energy Trading}.
\newblock {\em {IEEE Transactions on Smart Grid}}, 10(2):2026 -- 2035, 2019.

\bibitem{Nguyen-TSG17}
D.~H. Nguyen, T.~Narikiyo, and M.~Kawanishi.
\newblock Optimal demand response and real-time pricing by a sequential
  distributed consensus-based {ADMM} approach.
\newblock {\em IEEE Transactions on Smart Grid}, 63(6):1694--1700, 2018.

\bibitem{Ozawa18}
A.~Ozawa and Y.~Kudoh.
\newblock {Performance of residential fuel-cell-combined heat and power systems
  for various household types in Japan}.
\newblock {\em {International Journal of Hydrogen Energy}}, 43:15412--15422,
  2018.

\bibitem{Sorin19}
E.~Sorin, L.~Bobo, and P.~Pinson.
\newblock {Consensus-based approach to peer-to-peer Electricity Markets with
  Product Differentiation}.
\newblock {\em {IEEE Transactions on Power Systems}}, 34(2):994--1004, 2019.

\bibitem{Tran18}
H.~N. Tran, T.~Narikiyo, M.~Kawanishi, S.~Kikuchi, and S.~Takaba.
\newblock {Whole-day optimal operation of multiple combined heat and power
  systems by alternating direction method of multipliers and consensus theory}.
\newblock {\em {Energy Conversion and Management}}, 174:475--488, 2018.

\bibitem{Sousa19}
T.Sousa, T.~Soares, P.~Pinson, F.~Moret, T.~Baroche, and E.~Sorin.
\newblock {Peer-to-peer and community-based markets: A comprehensive review}.
\newblock {\em {Renewable and Sustainable Energy Reviews}}, 104:367--378, 2019.

\bibitem{Tushar19}
W.~Tushar, T.~K. Saha, C.~Yuen, T.~Morstyn, M.~D. McCulloch, H.~V. Poor, and
  K.~L.Wood.
\newblock {A motivational game-theoretic approach for peer-to-peer energy
  trading in the smart grid}.
\newblock {\em {Applied Energy}}, 243:10--20, 2019.

\bibitem{Tushar20}
W.~Tushar, T.~K. Saha, C.~Yuen, D.~Smith, and H.~V. Poor.
\newblock {Peer-to-Peer Trading in Electricity Networks: An Overview}.
\newblock {\em {IEEE Transactions on Smart Grid}}, 11(4):3185--3200, 2020.

\bibitem{Tushar18}
W.~Tushar, C.~Yuen, H.~Mohsenian-Rad, T.~Saha, H.~V. Poor, and K.~L. Wood.
\newblock {Transforming Energy Networks via Peer-to-Peer Energy Trading: The
  potential of game-theoretic approaches}.
\newblock {\em {IEEE Transactions on Power Systems}}, 35(4):90--111, 2018.

\bibitem{Wakui10}
T.~Wakui, R.~Yokoyama, and K.~Shimizu.
\newblock {Suitable operational strategy for power interchange operation using
  multiple residential SOFC (solid oxide fuel cell) cogeneration systems}.
\newblock {\em {Energy}}, 35:740--750, 2010.

\bibitem{Werth18}
A.~Werth, A.~Andre, D.~Kawamoto, T.~Morita, S.~Tajima, M.~Tokoro,
  D.~Yanagidaira, and K.~Tanaka.
\newblock {Peer-to-Peer Control System for DC Microgrids}.
\newblock {\em {IEEE Transactions on Smart Grid}}, 9(4):3667--3675, 2018.

\end{thebibliography}

\end{document}